\documentstyle[12pt,epsf]{article}

\begin{document}

\baselineskip=20pt

\renewcommand{\theequation}{\arabic{section}.\arabic{equation}}
\newfont{\elevenmib}{cmmib10 scaled\magstep1}
\newcommand{\YUKAWAmark}{\elevenmib
            Yukawa\hskip1mm Institute\, Kyoto}
\newcommand{\preprint}{\begin{flushleft} 
                \YUKAWAmark
                \end{flushleft}
                \vspace{-0.95cm}
                \begin{flushright}  \normalsize \sf
                YITP-96-52   \\
                quant-ph/9610021 \\
                September 1996
            \end{flushright}      }

\title{  \preprint \Large \bf \ \\ \ \\ 
         Hypergeometric States and Their Nonclassical Properties  }

\author{ Hong-Chen Fu\thanks{   On leave of absence from Institute
                                of Theoretical Physics, Northeast 
                                Normal University, Changchun 130024, 
                                P.R.China.
                                E-mail: hcfu@yukawa.kyoto-u.ac.jp }\ \
         and Ryu Sasaki \\  \\ {\normalsize \it
         Yukawa Institute for Theoretical Physics, Kyoto University,}\\
          {\normalsize \it Kyoto 606-01, Japan}}

\date{\ }
\maketitle

\ \\ 
\begin{abstract}
`Hypergeometric states', which are a one-parameter generalization of
binomial states of the single-mode quantized radiation field,
are introduced and their nonclassical
properties are investigated. Their limits to the binomial
states and to the coherent and number states are studied. The
ladder operator formulation of the hypergeometric states is found 
and the algebra involved turns out to be a
one-parameter deformation of $su(2)$ algebra. These states exhibit
highly nonclassical properties, like sub-Poissonian
character, antibunching and squeezing effects. The
quasiprobability distributions in phase space, namely the $Q$ and the
Wigner functions are studied in detail. These remarkable properties 
seem to suggest that the hypergeometric states deserve further 
attention from theoretical and applicational sides of quantum optics.
\\ \\
PACS numbers: 03.65.-w, 42.50.-p
\end{abstract}

\newpage

\section{Introduction}
\setcounter{equation}{0}

The number and the coherent states of the quantized radiation field
play important roles in quantum optics and are extensively studied
\cite{noch}. The binomial states (BS, also called intermediate
number-coherent states) introduced by Stoler, Saleh and Teich in
1985 \cite{stol}, interpolate between the {\it most nonclassical}
number states and the {\it most classical} coherent states, and
reduce to them in two different limits. Some of their properties
\cite{stol,leee,barr}, methods of generation \cite{stol,leee,datt},
as well as their interaction with atoms \cite{josh}, have been
investigated in the literature. The BS is defined as a linear
superposition of number states ($|n\rangle$, $n=0,1,\ldots$) in an 
$M+1$-dimensional subspace
\begin{equation}
  |\eta,\,M\rangle=\sum_{n=0}^{M}\beta_n^M(\eta)|n\rangle,
  \label{bs}
\end{equation}
where $\eta$ is a real parameter satisfying $0<\eta<1$ (probability),
and
\begin{equation}
  \beta_n^M(\eta)=\left[\left(\begin{array}{c}M\\n\end{array}\right)
  \eta^n (1-\eta)^{M-n}\right]^{1/2}.
  \label{bsdist}
\end{equation}
The name `binomial state' comes from the fact that their photon 
distribution $|\langle n|\eta,\,M\rangle|^2 \linebreak =
|\beta_n^M(\eta)|^2$ is
simply the binomial distribution with probability $\eta$.
In the two limits $\eta \to 1$ and  $\eta \to 0$
it reduces to number states:
\begin{equation}
   |1,\,M\rangle=|M\rangle, \qquad |0,\,M\rangle=|0\rangle.
   \label{numlim}
\end{equation}
In a different limit of $M\to \infty, \  \eta\to 0$ with
$\eta M =\alpha^2$ fixed ($\alpha$ real constant), $|\eta,\,M\rangle$
reduces to the  coherent states (not the most general
ones, only those with real amplitude $\alpha$), which corresponds to 
the Poisson distribution in probability theory\cite{Feller}.
It is well known that the binomial distribution tends to the Poisson 
distribution in the above limit\cite{Feller}.
The notion of BS was also generalized to the intermediate squeezed
states \cite{bas1} and the number-phase states \cite{bas2}, as
well as their $q$-deformation \cite{fann}. In particular, in a
previous paper \cite{fus4} we derived the ladder-operator form the BS and
on this basis we generalized the BS to the generalized BS which
possessed the number and squeezed states as  limits in the
framework of Lie algebra $su(2)$.

In the present paper we shall propose a one-parameter generalization
of the binomial states, the {\it hypergeometric states} (HGS), whose photon
distribution is the hypergeometric distribution in probability
theory \cite{Feller}. It is well known that the hypergeometric distribution tends to the
binomial distribution in certain limit. This leads to the reduction
of HGS to the BS in the same limit. Some mathematical
properties, such as the equivalent ladder operator definition,
related algebraic structures, will be formulated. It is interesting that
the algebraic structure is a well-investigated {\it generally
deformed oscillator algebra} \cite{gdo}, which reduces to the universal enveloping
algebra of Lie algebra $su(2)$, the algebraic structure characterizing the
binomial states. The nonclassical properties of the HGS,
the photon statistical properties, sub-Poissonian distribution, antibunching
effect and the squeezing effect will be investigated in detail.
Two well known quasi-probability distributions, the $Q$-function and Wigner
function will be evaluated to study the nonclassical properties. It 
will be shown that the HGS exhibit highly nonclassical behaviour.

\section{HGS and basic properties}
\setcounter{equation}{0}

\subsection{Definition}

The HGS is defined as a linear combination of number states in an
$(M+1)$-dimensional subspace
\begin{equation}
  |L,M,\eta\rangle = \sum_{n=0}^M H_n^M(\eta,L)
  |n\rangle, \label{hp}
\end{equation}
where the {\it probability} $\eta$ is a real parameter satisfying
$0<\eta <1$. $L$ is a real number satisfying
\begin{equation}
  L\geq \mbox{max}\{M\eta^{-1},M(1-\eta)^{-1}\}, \label{ldef}
\end{equation}
and
\begin{eqnarray}
  H_n^M(\eta,L)&=&\left[\left(\begin{array}{c}L\eta\\n\end{array}\right)
  \left(\begin{array}{c}L(1-\eta)\\M-n\end{array}\right)\right]^{
  \frac{1}{2}}
  \left(\begin{array}{c}L\\M\end{array}\right)^{-\frac{1}{2}},
  \label{hdef}\\
  \left(\begin{array}{c}\alpha\\n\end{array}\right)&=&
  \frac{\alpha(\alpha-1)\cdots(\alpha-n+1)}{n!},\ \ \ \ \
  \left(\begin{array}{c}\alpha\\0\end{array}\right)\equiv 1.
  \label{gbi}
\end{eqnarray}
Note that in Eq.(\ref{gbi}) the real number $\alpha$ is not
necessarily an integer.

The name of HGS comes from the fact that the photon distribution
$|\langle n|L,M,\eta\rangle|^2$
\begin{equation}
  |\langle n|L,M,\eta\rangle|^2=[H_n^M(\eta,L)]^2,
  \label{hgdist}
\end{equation}
is the hypergeometric distribution in probability theory
\cite{Feller}. (For a background, see the Appendix.) We remark that in the case of $M=1$, the HGS
$|L,1,\eta\rangle$ is $L$ independent and is equal to the binomial
state $|L,1,\eta\rangle\equiv |1,\eta\rangle $.

It is well known that the hypergeometric distribution tends to
the binomial distribution in the limit $L\to\infty$. So,
correspondingly, the HGS tends to the BS in this limit
\begin{equation}
   |L,M,\eta\rangle \stackrel{L\to\infty}{\longrightarrow}
   |M,\eta\rangle.
\end{equation}
This fact can be verified directly. Furthermore, the BS
go to the number states and coherent states, the latter corresponds to
the Poisson distribution, in certain limits. So the HGS reduce to
the number and coherent states in these limits:
\begin{equation}
   |L,M,\eta\rangle \stackrel{L\to\infty}{\longrightarrow}
   |M,\eta\rangle \longrightarrow \left\{ \begin{array}{ll}
   |M\rangle, & \mbox{when $\eta\to 1$},\\
   |0\rangle, & \mbox{when $\eta\to 0$}, \\
   |\alpha\rangle, & \mbox{when $M\to\infty$, $\eta\to 0$ with finite
                     $\eta M \equiv\alpha$.}
   \end{array} \right.
\end{equation}

It is easy to show that the HGS are normalized by using the following
identity
\begin{equation}
   \sum_{n=0}^M \left(\begin{array}{c}\alpha\\n\end{array}\right)
      \left(\begin{array}{c}\beta\\M-n\end{array}\right)=
      \left(\begin{array}{c}\alpha+\beta\\M\end{array}\right),
\end{equation}
which can be obtained by comparing the power series expansion of
$(1+t)^\alpha (1+t)^\beta=(1+t)^{\alpha+\beta}$.

\subsection{Ladder operator form}

In a previous paper \cite{fus4} we have shown that the BS
admit the ladder operator formulation, namely, they are
characterized by the
following eigenvalue equation
\begin{equation}
   [\sqrt{\eta}N+ \sqrt{1-\eta}J^+_M]|M,\eta\rangle=
   \sqrt{\eta}M|M,\eta\rangle, \label{bseq}
\end{equation}
where $J_M^+ \equiv \sqrt{M-N}\,a$ is the raising operator of the Lie
algebra $su(2)$ via its Holstein-Primakoff realization.
Hereafter $a^\dagger$ and $a$ are the creation and annihilation 
operators of the photon and $N=a^\dagger a$ is the number operator. 
So we naturally expect that the HGS satisfy a generalized eigenvalue
equation and the algebra involved is a deformation of $su(2)$.
To this end, we suppose that the HGS satisfy an eigenvalue equation
\begin{equation}
  [f(N)+g(N)a]|L,M,\eta\rangle=\lambda|L,M,\eta\rangle, \label{eq}
\end{equation}
in which $\lambda$ is the eigenvalue to be determined.
Inserting (\ref{hp}) into (\ref{eq}) and comparing the 
coefficients, we obtain
\begin{eqnarray}
  &&f(M)=\lambda, \label{eigenvalue}\\
  &&(\lambda-f(n))\left[\left(\begin{array}{c}L\eta\\n\end{array}\right)
     \left(\begin{array}{c}L(1-\eta)\\M-n\end{array}\right)\right]^{
     \frac{1}{2}}= \nonumber\\
  && \ \ \ \ \left[\left(\begin{array}{c}L\eta\\n+1\end{array}\right)
     \left(\begin{array}{c}L(1-\eta)\\M-n-1\end{array}\right)(n+1)
     \right]^{\frac{1}{2}}\,g(n),
     \ \ (0\leq n \leq M-1).  \label{ei2}
\end{eqnarray}
From (\ref{eigenvalue}) and (\ref{ei2}), we have
\begin{equation}
     f(M)-f(n)=\sqrt{\frac{L\eta-n}{L(1-\eta)-M+n+1}}\sqrt{M-n}
     \, g(n).
\end{equation}
Observe that in the right side of the above equation  $M$ and $n$
 appear in the form $M-n$. Requiring that Eq.(\ref{eq})
reduces to (\ref{bseq}) in the limit $L\to \infty$, we obtain
\begin{equation}
     f(N)=\sqrt{\eta}N, \ \ \ \
     g(N)=\sqrt{\eta}\left[\frac{L(1-\eta)-M+N+1}{L\eta-N}
     \right]^{\frac{1}{2}}\sqrt{M-N}. \label{fg}
\end{equation}
Substituting (\ref{fg}) into (\ref{eq}) we arrive at the
ladder operator form of HGS
\begin{equation}
     \left[\sqrt{\eta}N+\sqrt{\eta}\left(\frac{L(1-\eta)-M+N+1}
     {L\eta-N}\right)^{\frac{1}{2}}\,J_M^+\right]
     |L,M,\eta\rangle=\sqrt{\eta}M|L,M,\eta\rangle. \label{eifin}
\end{equation}
It is easy to see that (\ref{eifin}) reduces to (\ref{bseq})
in the limit of $L\to \infty$ for finite $M$ and $N$.

Here we would like to remark that the operator in the left
side of (\ref{eifin}) is an $(M+1)\times(M+1)$ matrix and
it generally has $M+1$ eigenvalues and eigenstates. The HGS is
only one eigenstate of the eigenvalue $\sqrt{\eta}M$.

Let us examine the algebraic structure involved in 
(\ref{eifin}). Define ${\cal A}(L,M)$ as an associative
algebra with generators
\begin{equation}
    N, \ \ \ \
    A_M^-=\left(\frac{\eta}{1-\eta}\right)^{\frac{1}{2}}
          \left(\frac{L(1-\eta)-M+N+1}{L\eta-N}\right)^{
          \frac{1}{2}}\,J_M^+,\ \ \ \
    A_M^+=(A_M^-)^{\dagger}.
\end{equation}
Then it is easy to verify that these operators satisfy the following
commutation relations
\begin{equation}
   [N, A_M^{\pm}]=\pm A_M^{\pm}, \ \ \ \
   A_M^+ A_M^- = F(N),\ \ \ \
   A_M^- A_M^+ = F(N+1),\ \ \ \
\end{equation}
where the function $F(N)$
\begin{equation}
   F(N)=\frac{\eta[L(1-\eta)-M+N]N(M-N+1)}
        {(1-\eta)(L\eta-N+2)},
\end{equation}
is non-negative for $0\leq N\leq M$. This algebra ${\cal A}(L,M)$
is nothing but the
{\it generally deformed oscillator algebra} with the structure function
$F(N)$ \cite{gdo}. In terms of the generators of ${\cal A}(L,M)$,
Eq.(\ref{eifin}) can be rewritten as
\begin{equation}
   \left[\sqrt{\eta}N+\sqrt{1-\eta}A_M^- \right]|L,M,\eta\rangle
   =\sqrt{\eta}M|L,M,\eta\rangle.
\end{equation}

It is interesting that this algebra ${\cal A}(L,M)$ is an
$L$-deformation of $su(2)$ in the sense that it contracts to
the universal enveloping algebra of the Lie algebra $su(2)$ in the
limit $L\to\infty$ with $M$ and $N$ finite
\begin{equation}
   A_M^{\pm}\stackrel{L\to\infty}{\longrightarrow}
   J_M^{\mp}.
\end{equation}
This means that the ladder operator form reduces to that of the
BS.

\section{Nonclassical Properties}
\setcounter{equation}{0}

In this section we turn to the nonclassical properties of the
HGS.

\subsection{Mean photon number and fluctuation}

The mean photon number in the HGS is obtained as
\begin{equation}
   \langle N\rangle=\langle L,M,\eta|N|L,M,\eta\rangle=M\eta.
   \label{meanen}
\end{equation}
It is interesting that it is independent of $L$ and therefore it is exactly
same as that of the BS. However, the mean
value of $N^2$ depends on $L$
\begin{equation}
   \langle N^2\rangle=M\eta\frac{L+L\eta M-L\eta-M}
   {L-1}.
\end{equation}
Then the fluctuation of the photon number is
\begin{equation}
   \langle \Delta N^2\rangle \equiv \langle N^2\rangle-
   \langle N\rangle^2=\eta(1-\eta)M\frac{L-M}{L-1}
   =\langle \Delta N^2\rangle_{BS}\frac{L-M}{L-1},
\end{equation}
where the $\langle \Delta N^2\rangle_{BS}\equiv \eta(1-\eta)M$
is the corresponding fluctuation of the BS. We find that this
fluctuation is always weaker than that of the BS
since the factor $W(L,M)=(L-M)/(L-1)$ is always smaller than 1
except for $M=1$ and the limit $L\to \infty$. Let us go into
some detail of the factor $W(L,M)$,  which is referred to as
the {\it weakening factor}. We shall see that this factor lies
between one half and 1
\begin{equation}
   \frac{1}{2}<W(L,M)<1.
\end{equation}
In fact, as a function of $L$ for fixed $M$, $W(L,M)$ is an
increasing
function and the smallest $W(L,M)$ corresponds to the
smallest $L$, which is $2M$ for $\eta=0.5$. In this case
the weakening factor is rewritten as $W(M)=M/(2M-1)$ which
is always larger than $1/2$.

So, in comparison with the BS, the fluctuation
is reduced in the HGS. For  large $M$ with $\eta=0.5$,
the fluctuation is only about half as much as the binomial
states. This is an important feature of the HGS.

\subsection{Sub-Poissonian distribution}

Let us introduce 
Mandel's $Q$ parameter \cite{Man} defined by
\begin{equation}
   Q=\frac{\langle \Delta N^2\rangle - \langle N\rangle}
   {\langle N \rangle},
\end{equation}
which measures the deviation from the Poisson distribution 
(the coherent state, $Q=0$).
If $Q<0\ (>0)$, the field is called sub(super)-Poissonian,
respectively. For the HGS, it is easy to
find that
\begin{equation}
   Q=(1-\eta)W(L,M)-1,
\end{equation}
which is generally negative since $1-\eta<1$ and $W(L,M)< 1$,
namely, the field on the HGS is sub-Poissonian. Exceptions
are the coherent state limit ($L\to\infty,\ M\to \infty,\ \eta
\to 0$ with $\eta M=\alpha^2$ finite) and the vacuum state limit
($L\to\infty,\ \eta\to 0$). The extreme case is $Q=-1$ since
$(1-\eta)W(L,M)$ is always positive. In the case $\eta \to 1$
and $L=(1-\eta)^{-1}M\to \infty$, namely, on the number states,
the extreme case occurs.

\subsection{Antibunching effect}

We say the field is antibunched if the second-order correlation
function $g^{(2)}(0)$ satisfies \cite{WaMi}
\begin{equation}
   g^{(2)}(0)=\frac{\langle a^{\dagger}a^{\dagger}a a\rangle}
   {\langle a^{\dagger}a\rangle^2} =
   \frac{\langle N^2-N\rangle}{\langle N\rangle^2}<1.  \label{an}
\end{equation}
For the HGS 
\begin{equation}
   g^{(2)}(0)=\frac{M-1}{M}\,\frac{L\eta-1}{L\eta-\eta}
\end{equation}
which always satisfies the condition (\ref{an}). So the HGS is
antibunched. In fact, the occurrence of antibunching effect
and sub-Poissonian are concomitant for single mode and time
independent fields.

In the binomial state limit $L\to\infty$, $g^{(2)}(0)$ reduces
to the second-order correlation function $g^{(2)}(0)_{BS}$ of
the BS
\begin{equation}
   g^{(2)}(0)\stackrel{L\to\infty}{\longrightarrow}
   g^{(2)}(0)_{BS}=\frac{M-1}{M}.
\end{equation}
Since the factor $(L\eta-1)/(L\eta-\eta)<1$, the HGS are more
strongly antibunched than the BS.

\subsection{Squeezing effect [1,4]}

Define the quadrature operators $x$ (coordinate) and $p$
(momentum) by
\begin{equation}
   x=\frac{1}{\sqrt{2}}(a^{\dagger}+a), \ \ \ \
   p=\frac{i}{\sqrt{2}}(a^{\dagger}-a).
   \label{coomom}
\end{equation}
Then their variances
\begin{equation}
   \langle \Delta x^2\rangle=\langle x^2\rangle-\langle x\rangle^2, \ \ \ \
   \langle \Delta p^2\rangle=\langle p^2\rangle-\langle p\rangle^2, \ \ \ \
\end{equation}
obey the Heisenberg's uncertainty relation
\begin{equation}
   \langle\Delta x^2\rangle\langle\Delta p^2\rangle \geq \frac{1}{4}.
\end{equation}
If one of the $\langle\Delta x^2\rangle$ and $\langle\Delta p^2
\rangle$ is smaller than 1/2, the squeezing occurs.
For convenience, we define the squeezing indices
\begin{equation}
   S_x=\frac{\langle\Delta x^2\rangle -1/2}{1/2},\ \ \ \
   S_p=\frac{\langle\Delta p^2\rangle -1/2}{1/2}.
\end{equation}
If $S_x<0$ ($S_p<0$), there is  squeezing in the quadrature
$x$ ($p$). Now let us evaluate these indices.

It is easy to derive that
\begin{eqnarray}
  a^n |L,M,\eta\rangle &=&\left(\begin{array}{c}L\\M\end{array}
  \right)^{-1/2}\sqrt{L\eta(L\eta-1)\cdots (L\eta-n+1)}\nonumber \\
  &&\times\sum_{k=0}^{M-n}\left[\left(\begin{array}{c}L\eta-n\\k\end{array}
  \right)\left(\begin{array}{c}L(1-\eta)\\M-n-k\end{array}
  \right)\right]^{1/2}|k\rangle
\end{eqnarray}
for $n\leq M$ and $a^n|L,M,\eta\rangle=0$ for $n>M$. In particular,
for $n=1,2$, we write
\begin{equation}
   a|L,M,\eta\rangle=\sum_{k=0}^{M-1} \bar{H}_k |k\rangle, \ \ \ \
   a^2|L,M,\eta\rangle=\sum_{k=0}^{M-2} \tilde{H}_k |k\rangle,
\end{equation}
where
\begin{eqnarray}
   \bar{H}_k&=&\sqrt{L\eta}\left(\begin{array}{c}L\\M
   \end{array}\right)^{-\frac{1}{2}}\left[\left(\begin{array}{c}L\eta-1\\
   n\end{array}\right)\left(\begin{array}{c}L(1-\eta)\\M-1-k
   \end{array}\right)\right]^{\frac{1}{2}}, \ \
   (0\leq k\leq M-1), \nonumber \\
   \tilde{H}_k&=&\sqrt{L\eta(L\eta-1)}\left(\begin{array}{c}L\\M
   \end{array}\right)^{-\frac{1}{2}}\left[\left(\begin{array}{c}L\eta-2\\
   n\end{array}\right)\left(\begin{array}{c}L(1-\eta)\\M-2-k
   \end{array}\right)\right]^{\frac{1}{2}}, \nonumber\\
   && (0\leq k\leq M-2).
\end{eqnarray}

In terms of $\bar{H}_n$ and $\tilde{H}_n$, we can obtain the
squeezing indices as
\begin{eqnarray}
   &&S_x=2\sum_{n=0}^{M-2}H_n\tilde{H}_n + 2M\eta-4\left[
       \sum_{n=0}^{M-1}H_n\bar{H}_n\right]^2, \\
   &&S_p=2M\eta-\sum_{n=0}^{M-2}H_n\tilde{H}_n,
\end{eqnarray}
in which $H_n\equiv H_n^M(\eta,L)$ in (\ref{hdef}). We have also 
suppressed the $\eta$ and $L$ dependence in $\bar{H}_n$ and $\tilde{H}_n$.

Figures 1 and 2 are plots showing how the $S_x$ depends on the
parameter $L$ and $\eta$. In each case,  different values of
$M$ (5 and 50) are chosen. From these plots we find that:

(1). Dependence on $L$ (Fig.1): When $\eta$ or $1-\eta$ is small, 
$|S_x|$ is always larger than that of the BS. The
smaller  $L$, the larger  $|S_x|$. However, the difference
from the BS is also small. This is understandable
since in this case, $L$ must be much larger than $M$ due to
the condition (\ref{ldef}), and therefore the HGS are 
close to the BS. When $\eta$ is around 0.5,
the $L$ can be closest to $M$ (two times), and the HGS are
far different from the BS. In particular,
when $L$ is small (close to $2M$),  $S_x$ changes
drastically in comparison with those of the BS.
In general the squeezing is great for large $M$ (50, see
Fig.1(b)) and it
decreases for  small $M$ (5, see Fig.1(a)).

(2). Dependence on $\eta$ (Fig.2): We have chosen $\eta=0.25
\sim 0.75$ for $M=5$ and $\eta=0.05\sim 0.95$ for $M=50$.
Similar to the BS, the squeezing increases
as $\eta$ increases  to a maximal point, and then it
decreases rapidly. This similarity is easy to understand.
In order to have a wide range of $\eta$,  $L$ is much
larger than $M$. In this case, the HGS are  closer
to the BS. We find that the HGS exhibits
stronger squeezing than the BS.

(3). Dependence on $M$: From Fig.1 we find that when $\eta$
is around 0.5, and for small $L$, the squeezing is weaken
for small $M$ (5) than large $M$ (50). From Fig.2
we conclude that  large $M$ has  wider and stronger squeezing range
of $\eta$ than the small $M$. The HGS has  wider
squeezing range of $\eta$ than the BS. This
can be seen from Fig.1 with $\eta=0.72$ (a) and $\eta=0.923$ (b).

\section{$Q$ and Wigner functions}
\setcounter{equation}{0}

The  quasi-probability distributions \cite{HiOc} in the coherent state
basis turn out to be useful measures for studying
the nonclassical features of the radiation field. In this section
we shall study the $Q$ and Wigner functions.

\subsection{$Q$-function}

We start with the $Q(\beta)$ function
\begin{equation}
   Q(\beta)=\frac{1}{\pi}|\langle \beta|L,M,\eta\rangle|^2,
   \label{qfunc}
\end{equation}
where $|\beta\rangle$ is the coherent states of the field.
Substituting the HGS into (\ref{qfunc}) we obtain the
$Q$-function as follows
\begin{equation}
   Q(\beta)=\frac{\exp(-|\beta|^2)}{\pi}
            \left|\sum_{n=0}^M H_n^M(\eta,L)
            \frac{\beta^n}{\sqrt{n!}}\right|^2.
\end{equation}
Here $\beta$ is a complex $c$-number $\beta=x+iy$, with
$(x,y\equiv p)$ corresponding to the two quadrature
operators $x$ and $p$ in (\ref{coomom}).

Now we would like to  investigate numerically the changes
of $Q$-function for different $L, M$ and $\eta$. Fig.3
are plots of $Q$-function of HGS for different $L$, for
fixed $M=5, \eta=0.5$. When $L\to \infty$, the HGS is in
fact a binomial state and its $Q$-function is shown in
Fig.3(d) (see also \cite{barr}). Then we choose  finite $L$
values 40 (Fig.3(c)),
20 (Fig.3(b)) and 10 (Fig.3(a)) (note that 10 is the smallest
allowed value of $L$ for $M=5, \eta=0.5$). We can see  clear
deformation of the $Q$-function. At first sight this deformation
pattern appears similar to that of $Q$ function with respect to
$\eta$ given in Fig.4 of \cite{barr}.  However, they are
essentially different: Increase in $\eta$ brings the gain of
the energy or the mean photon number as given (\ref{meanen}),
while the changes of $L$ does not correspond to any change
of the mean energy due to Eq.(\ref{meanen}). Fig.3(e) is the
$Q$-function of HGS for $\eta=0.9$, $L=50$ and $M=5$. This
$Q$-function is almost the same as that of BS (see Fig.4(c)
in Ref.\cite{barr}), as expected.

From the $Q$-functions we can also study the squeezing properties
by examining the deformation of their contours. As
before we pay our attention to the case $\eta=0.5$ and explain
the drastic changes of squeezing for small $L$. Fig.4 are the plots
of contours of $Q$-function for $M=5$, $\eta=0.5$ and $L=10$,
28 and $\infty$ (binomial states). We find that, when we decrease
$L$, the contour is first squeezed (Fig.4(b)) in the $x$
direction until a maximum squeezing is reached. Then the contour
deforms to the shape of an {\it ear}, which occupies a  wider range in the
$x$ direction and the squeezing is reduced (Fig.4(a)). From this
approach we can also explain the drastic increase of
squeezing for $M=50$ and $\eta=0.5$. In fact, when $L$ becomes
smaller, the shape of the contour is compressed, which corresponds
to the strong squeezing. And this change continues until the
smallest value of $L$ allowed for $\eta=0.5$, ie. $L=100$. 
However, in contrast to the case $M=5$,
the case $M=50$ does not give rise to the shape of an {\it ear} 
(Fig.4(d)).

\subsection{Wigner function}

The Wigner function in the series form is defined as \cite{wigser}
\begin{equation}
   W(\beta)=\frac{2}{\pi}\sum_{k=0}^\infty
   (-1)^k \langle \beta,k|\rho|\beta,k\rangle,
\end{equation}
where $|\beta,k\rangle=D(\beta)|k\rangle\equiv \exp(\beta a^{\dagger}
-\beta^* a) |k\rangle$, $\rho$ is the density matrix (projector on 
the HGS) and takes the form
\begin{equation}
   \rho=|L,M,\eta\rangle\langle L,M,\eta|
\end{equation}
for the case in hand. It is easy to compute that
\begin{equation}
   W(\beta)=\frac{2}{\pi}\sum_{k=0}^{\infty}(-1)^k \left|
   \sum_{n=0}^M H_n^M(\eta,L)\chi_{nk}(\beta)\right|^2.
\end{equation}
Here $\chi_{nk}(\beta)=\langle n|D(\beta)|k\rangle$ are
given by \cite{1333}
\begin{equation}
   \chi_{nk}(\beta)=\left\{\begin{array}{ll}
     \displaystyle{\sqrt{\frac{k!}{n!}}}
     \exp(-|\beta|^2/2) \beta^{n-k}
     {\cal L}_k^{n-k}(|\beta|^2)  & \mbox{if }n\geq k, \\
     \displaystyle{\sqrt{\frac{n!}{k!}}}
     \exp(-|\beta|^2/2) (\beta^*)^{k-n}
     {\cal L}_n^{k-n}(|\beta|^2)  & \mbox{if }n\leq k,
     \end{array}
     \right.
\end{equation}
where ${\cal L}_n^\alpha(|\beta|^2)$ are the generalized
Laguerre polynomials.

Since the case $M=1$ is simply the binomial state and its
Wigner functions have been investigated in detail in
Ref.\cite{barr}, we here consider the simplest nontrivial
case: $M=2$. Fig.5 are some plots of the Wigner function of
HGS for $M=2$ and different $\eta$: (a) $\eta=0.2$ (b)
$\eta=0.5$, (c) $\eta=0.9$ and (d) the number state
$|2\rangle$ ($\eta\to 1$ and $L\to \infty$). In each case
we choose  the smallest possible value of $L$ for given  $\eta$ and 
$M$ to see the maximal contrast with the BS. We note that
the case $\eta\to 0$ is just the vacuum state and its
Wigner function is simply the Gaussian centered at the origin.
As $\eta$ increases from 0, this Gaussian distribution continuously deforms
to the Wigner function of $|2\rangle$, as shown in Fig.5.
From $\eta=0.2$, the negative parts of the Wigner functions are
very clearly visible and this signifies the nonclassical properties.

Fig.6 are two plots of the Wigner function of the BS
for $M=2$ and (${\rm b}^\prime$) $\eta=0.5$ and (${\rm c}^\prime$) 
$\eta=0.9$. Comparing them
with those of HGS, namely, Fig.5 (b,c), we find that in the
case $\eta=0.5$ the Winger distributions of HGS and BS are
markedly different: distribution of HGS has two negative
peaks while the BS has only one. However, in the case of
$\eta=0.9$, two distributions are almost the same. This is
understandable since for $\eta=0.9$, $L$ must be very big and 
the HGS are very close to the BS.

\section{Conclusion}

We have shown various properties of the {\it hypergeometric 
states}. The relationship with the BS is clarified together 
with the coherent state and the number state limits. The 
ladder operator formulation gives an algebraic characterization 
of the HGS based on the generally deformed oscillator algebra.
The  salient statistical properties of the HGS such as the 
sub-Poissonian character, the anti-bunching effect and the 
squeezing effects are investigated for a wide range of the 
parameters. The nonclassical features of the HGS for certain 
parameter ranges are demonstrated in terms of the 
quasiprobability distributions, the Q-function and the Wigner 
function. On account of these remarkable properties we are 
tempted to think that the HGS  play an important role in 
quantum optics. Surely they deserve further investigation 
including the method of generation.


\section*{Acknowledgments}

This work is supported partially by the grant-in-aid for
Scientific Research, Priority Area 231 ``Infinite Analysis'',
Japan Ministry of Education. H.\,C.\,F is grateful to Japan
Society for Promotion of Science (JSPS) for the fellowship.
He is also supported in part by the National Science
Foundation of China.

\section*{Appendix.\ Hypergeometric Distribution}

Consider a pot containing $L_1$ red and $L_2$ black balls.
A group of $M$ balls is chosen randomly. Then the probability $q_n$ 
that the chosen group contains exactly $n$ red balls is given by
the hypergeometric distribution
$$
q_n={{L_1\choose n}{L_2\choose M-n}{L\choose M}^{-1}},\quad L=L_1+L_2.
$$
It is easy to see $q_n=|H_n^M(\eta,L)|^2$ (\ref{hgdist}) for 
$\eta=L_1/L$. The name is explained by the fact that the generating 
function of the above distribution can be expressed in terms of the
hypergeometric functions \cite{KenSt}. Obviously the above distribution
tends to the binomial distribution with probability $\eta$ in the large 
$L$ limit.


\newpage
%
\begin{figure} 
\centerline{\epsfxsize=9cm
\epsfbox{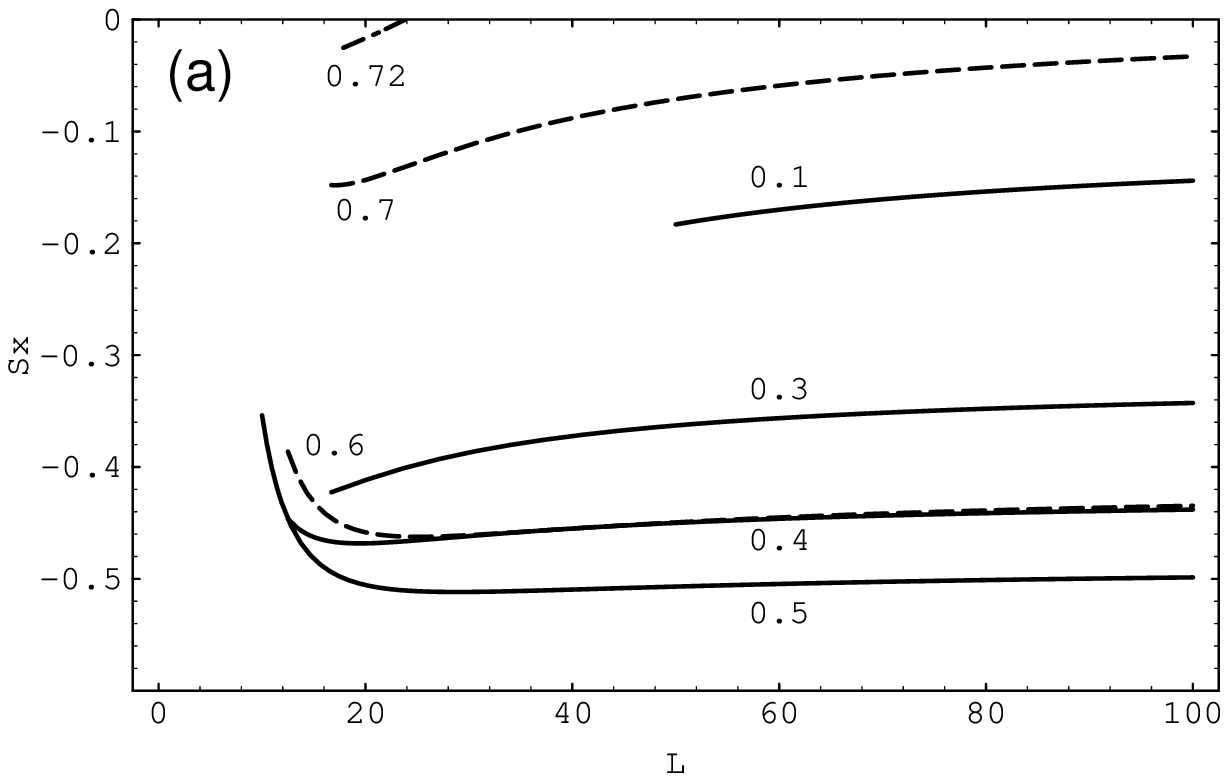}
\epsfxsize=9cm 
\epsfbox{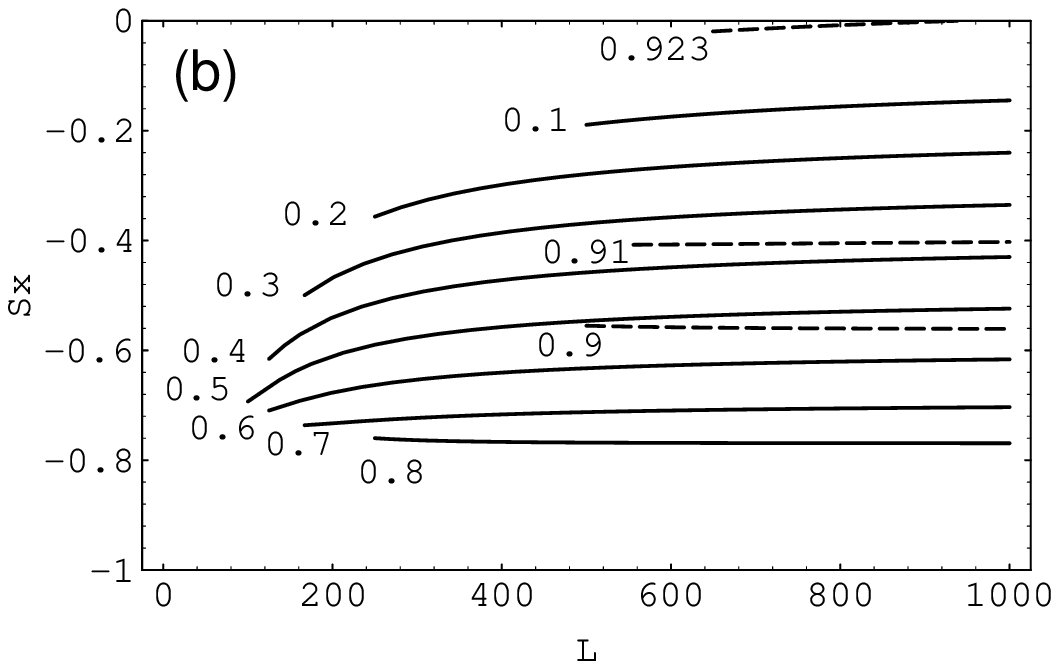}}
\caption{ Squeezing index $S_x$ of HGS as a function of $L$,
          for (a) $M=5$ and (b) $M=50$. The $\eta$ values are
          indicated in the figures. }
\end{figure}
%
\begin{figure}
\centerline{\epsfxsize=9cm
\epsfbox{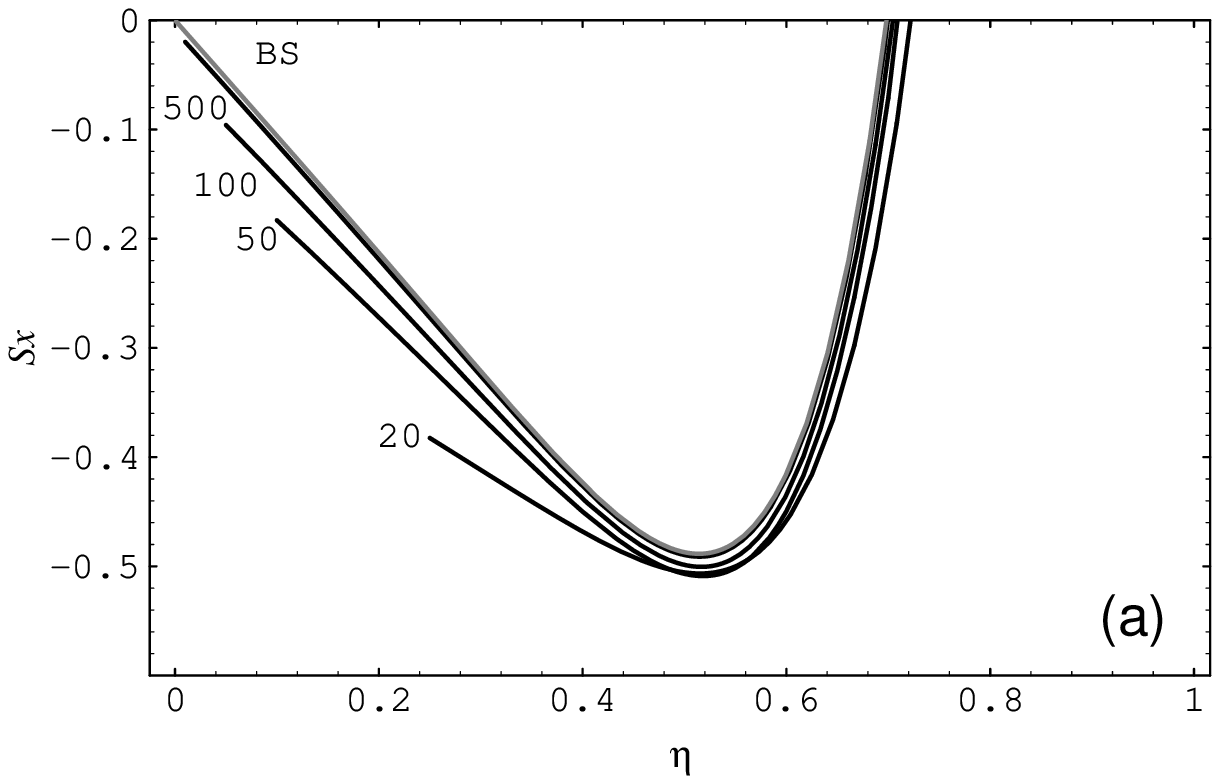}
\epsfxsize=9cm
\epsfbox{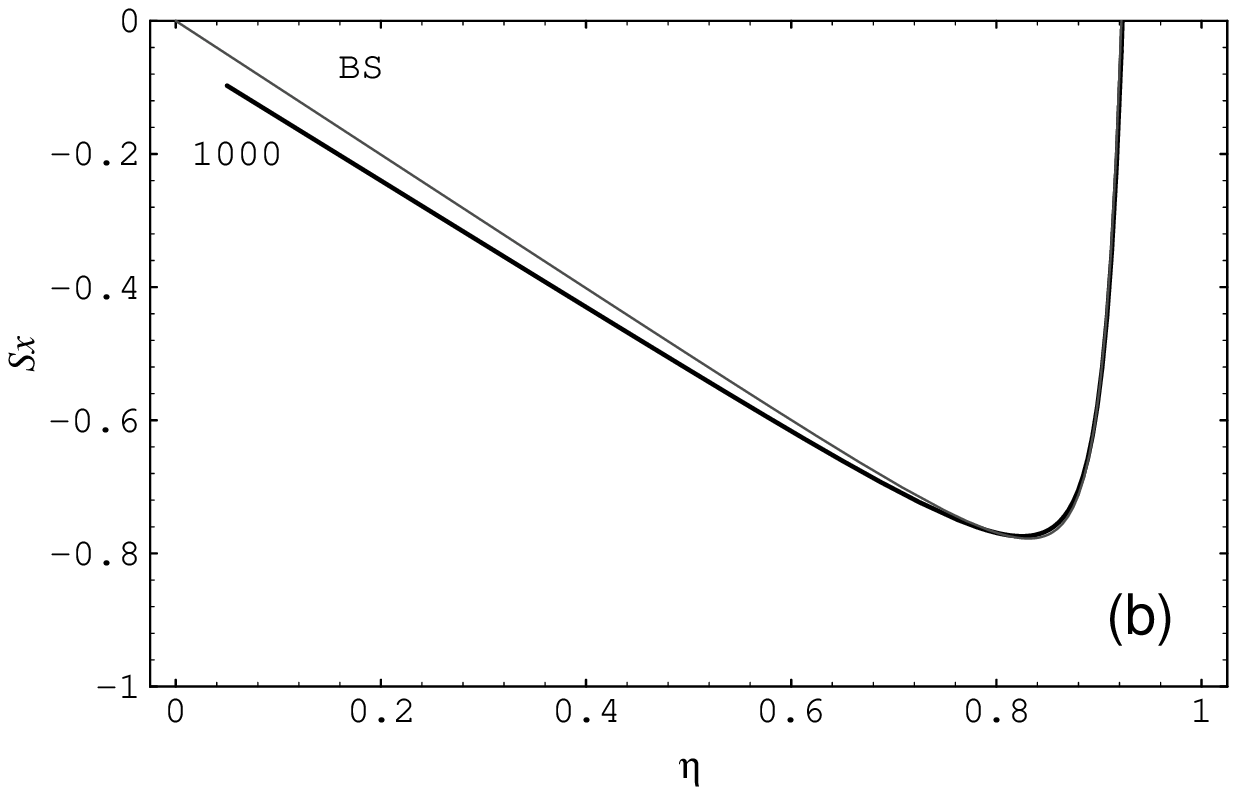}}
\caption{Squeezing index $S_x$ of HGS as a function of $\eta$ and for
different $M$ values: (a) $M=5$ and (b) $M=50$. The $L$ values
are shown in the figures.}
\end{figure}

\begin{figure}
\centerline{
\epsfxsize=8cm
\epsfbox{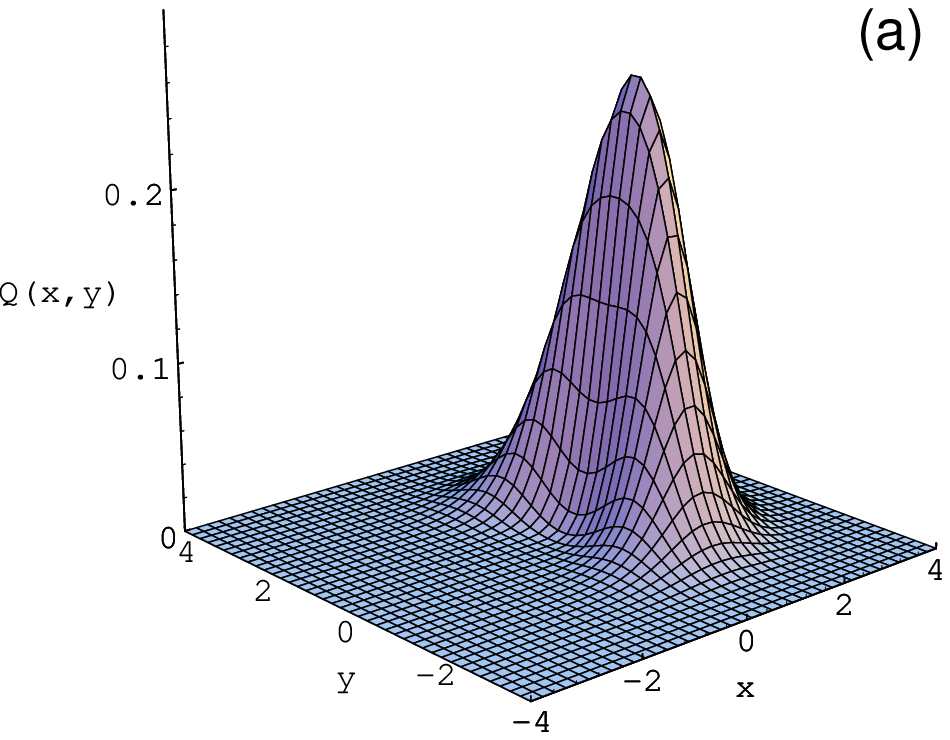}\hspace{1cm}
\epsfxsize=8cm
\epsfbox{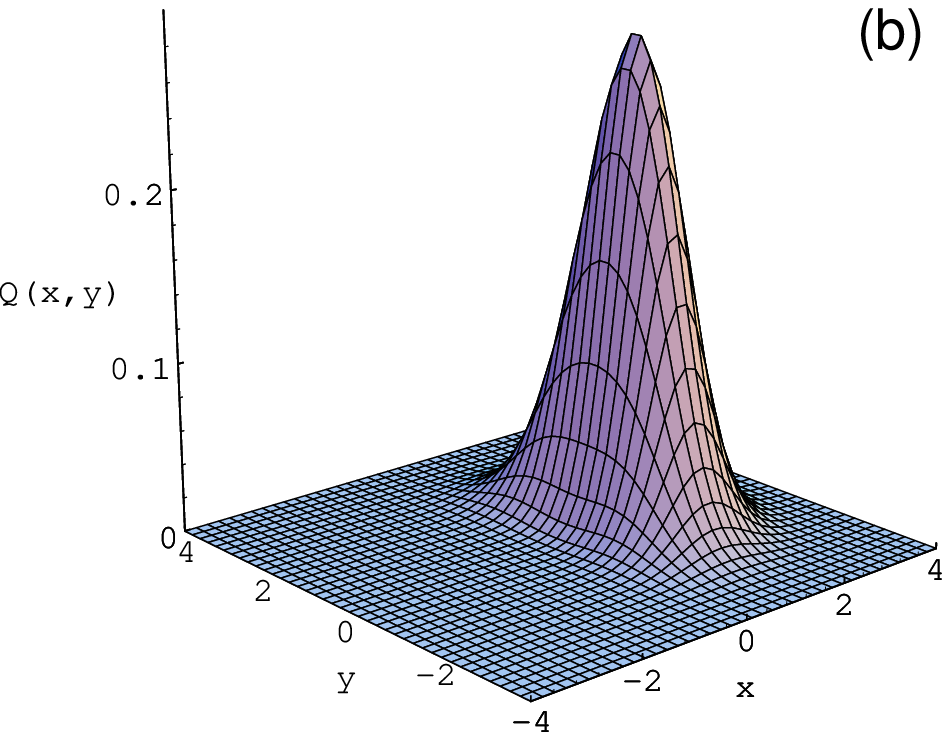}}
\centerline{	
\epsfxsize=8cm
\epsfbox{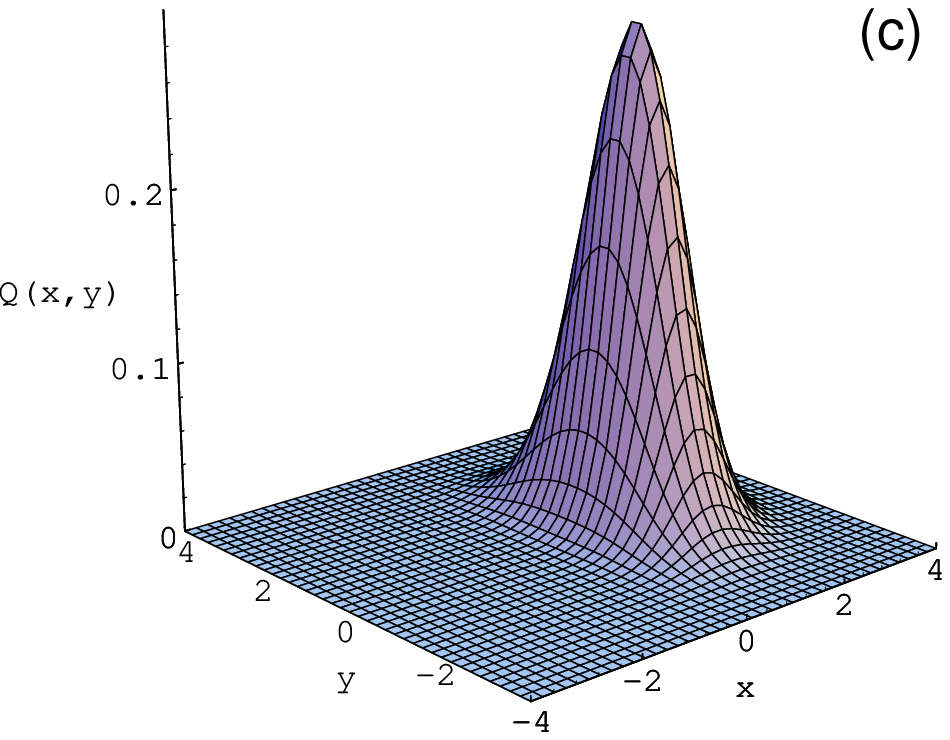}}
\centerline{
\epsfxsize=8cm
\epsfbox{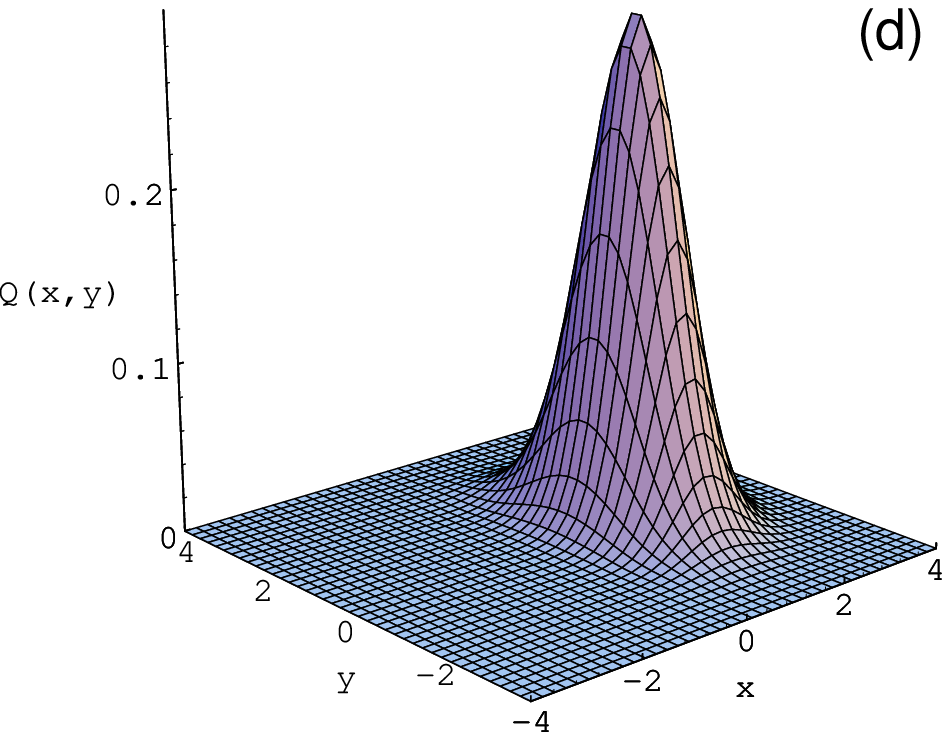}\hspace{1cm}
\epsfxsize=8cm
\epsfbox{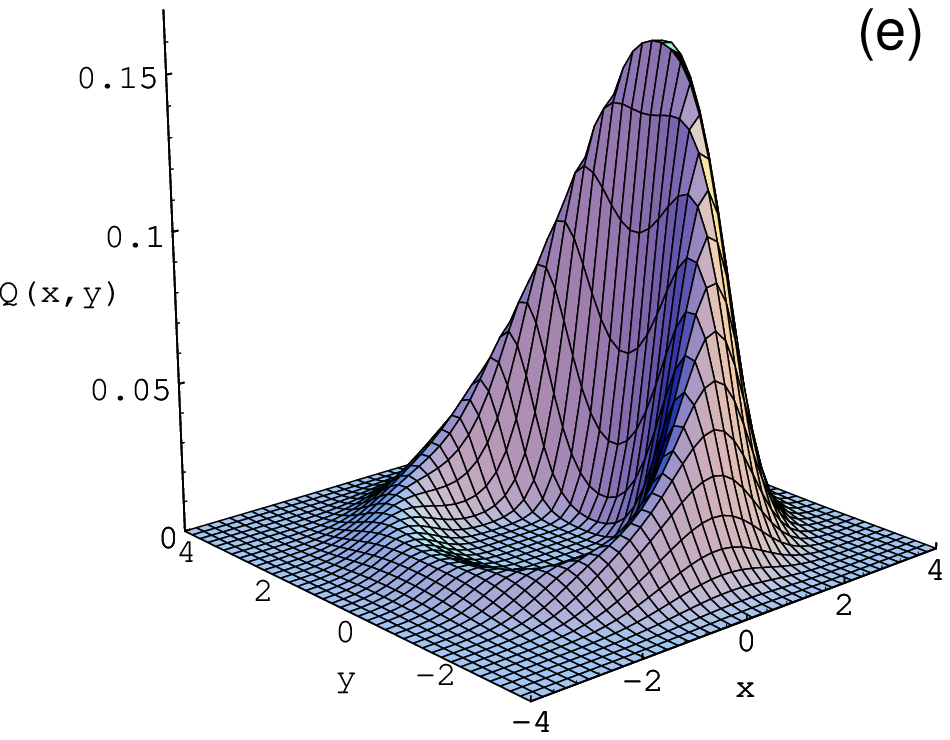}}
\caption{ $Q(\beta)$-functions of HGS. $\beta=x+iy$.\ (a) $\eta=0.5,\ L=10$,
(b) $\eta=0.5,\ L=20$, (c) $\eta=0.5,\ L=40$, (d)
$\eta=0.5,\ L\to\infty$ (binomial state). Those four
figures show how $Q$-function depends on $L$. Case
(e) corresponds to $\eta=0.9$ and $L=50$. In all the 
cases $M=5$.}
\end{figure}

%
\begin{figure}
\centerline{\epsfxsize=7cm 
\epsfbox{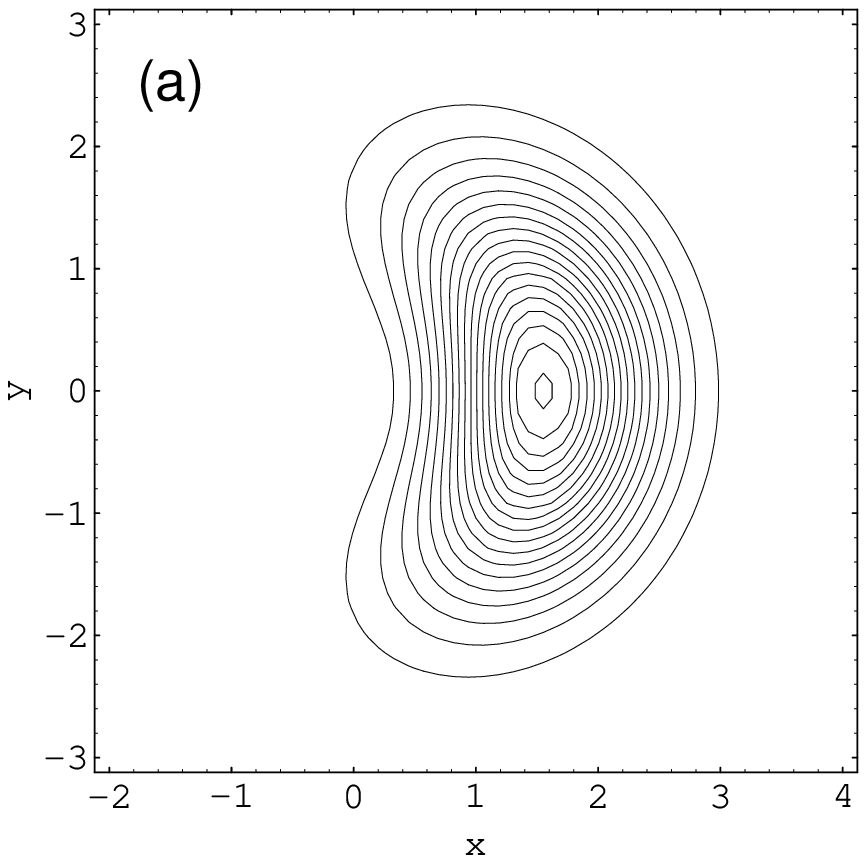}\hspace{1cm}
\epsfxsize=7cm 
\epsfbox{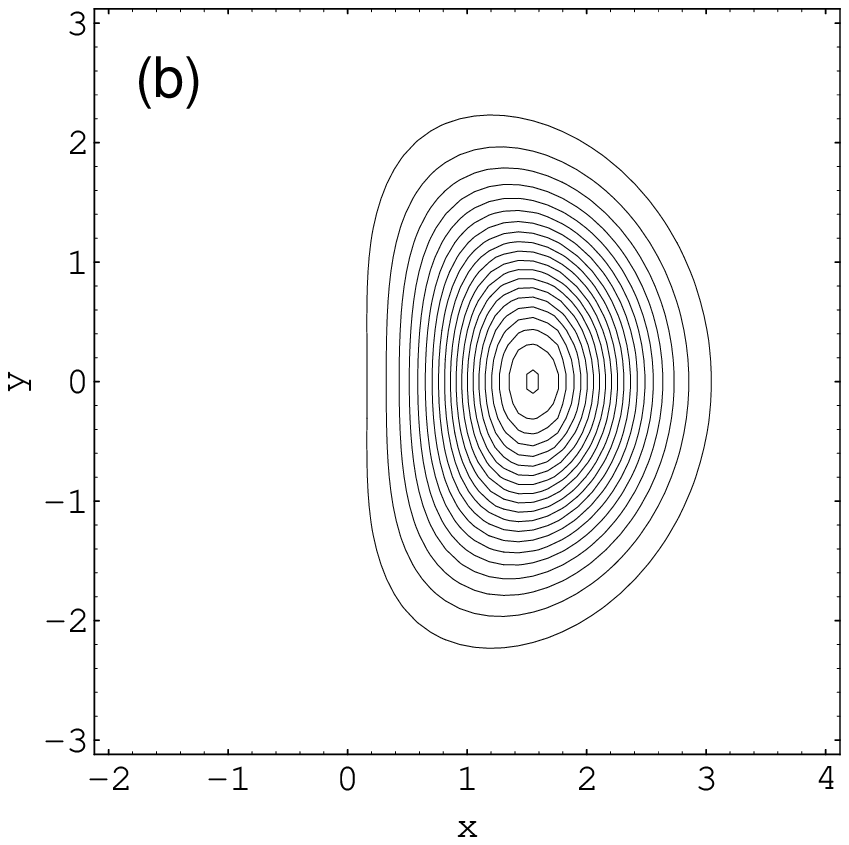}}
\centerline{\epsfxsize=7cm
\epsfbox{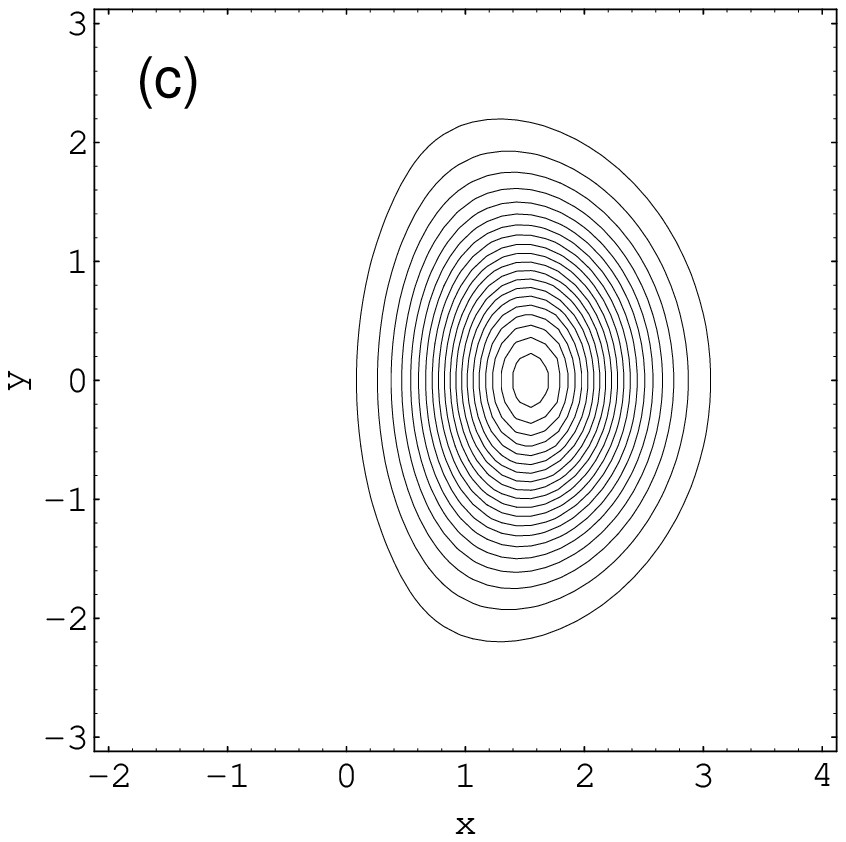}\hspace{1cm}
\epsfxsize=7cm
\epsfbox{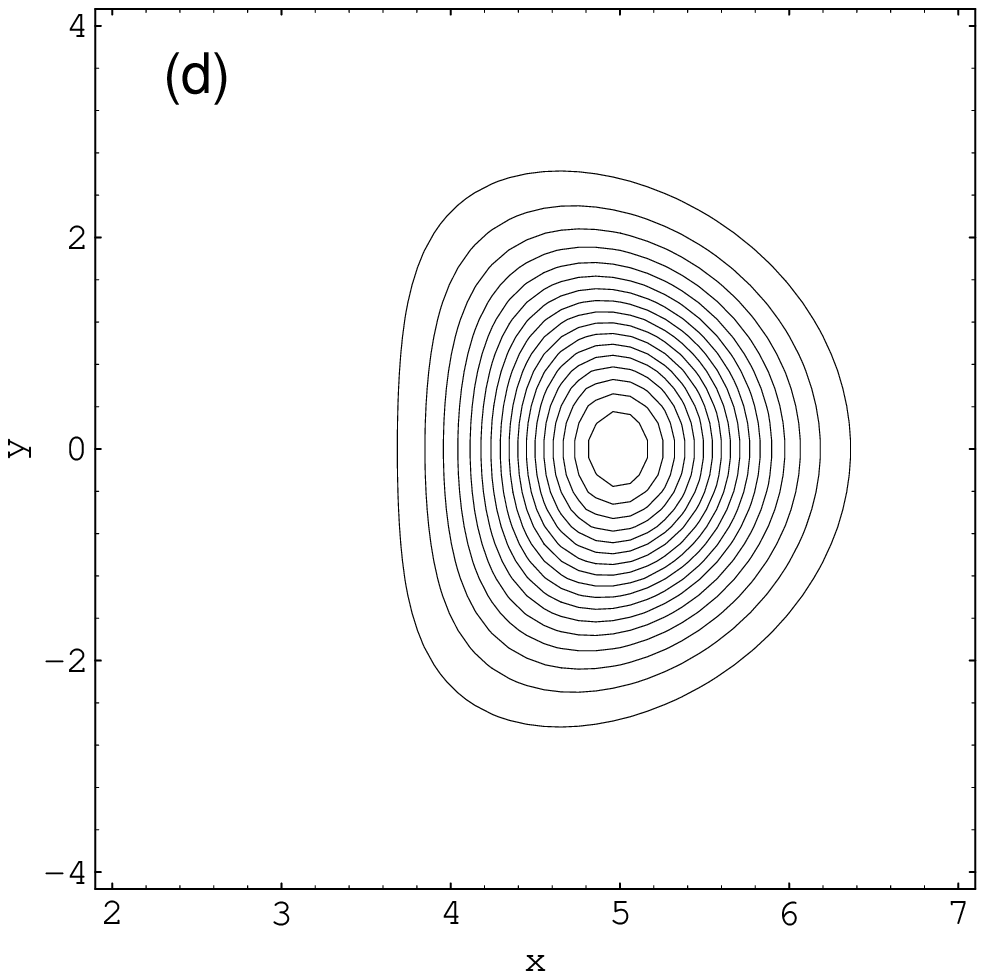}}
\caption{Contours of $Q(\beta)$-functions of HGS for $M=5$,
$\eta=0.5$ and (a) $L=10$, (b) $L=28$ and (c) $L=\infty$
(Binomial state); (d) $M=50$, $\eta=0.5$ and $L=100$.
$\beta=x+iy$.\  }
\end{figure}

\begin{figure}
\centerline{\epsfxsize=8cm 
\epsfbox{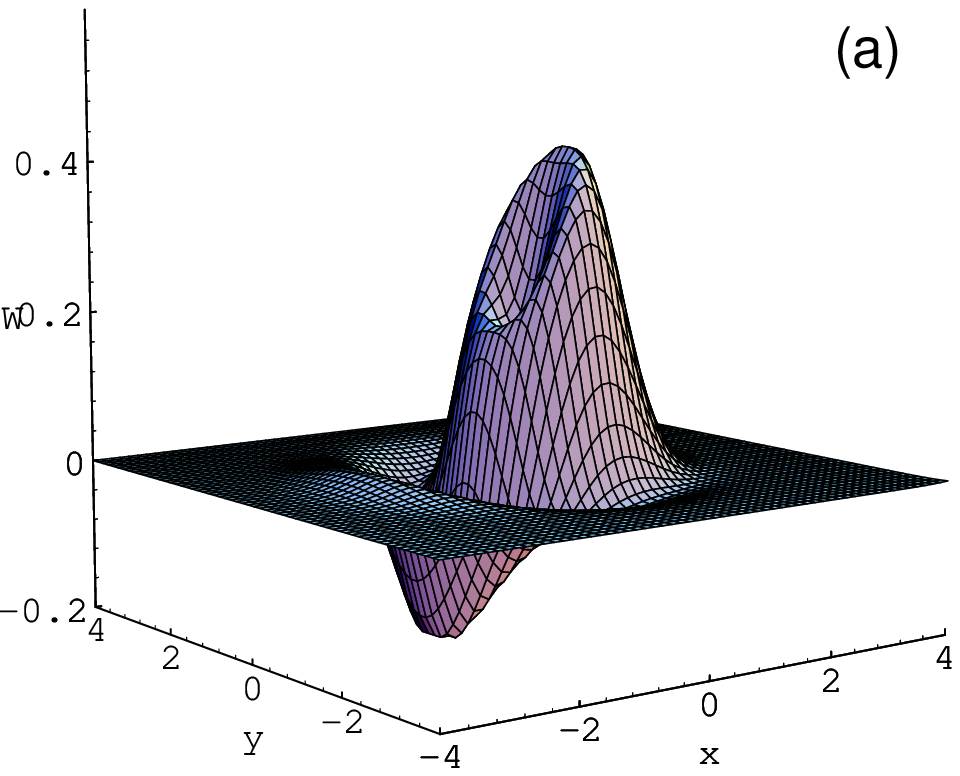}\hspace{1cm}
\epsfxsize=8cm 
\epsfbox{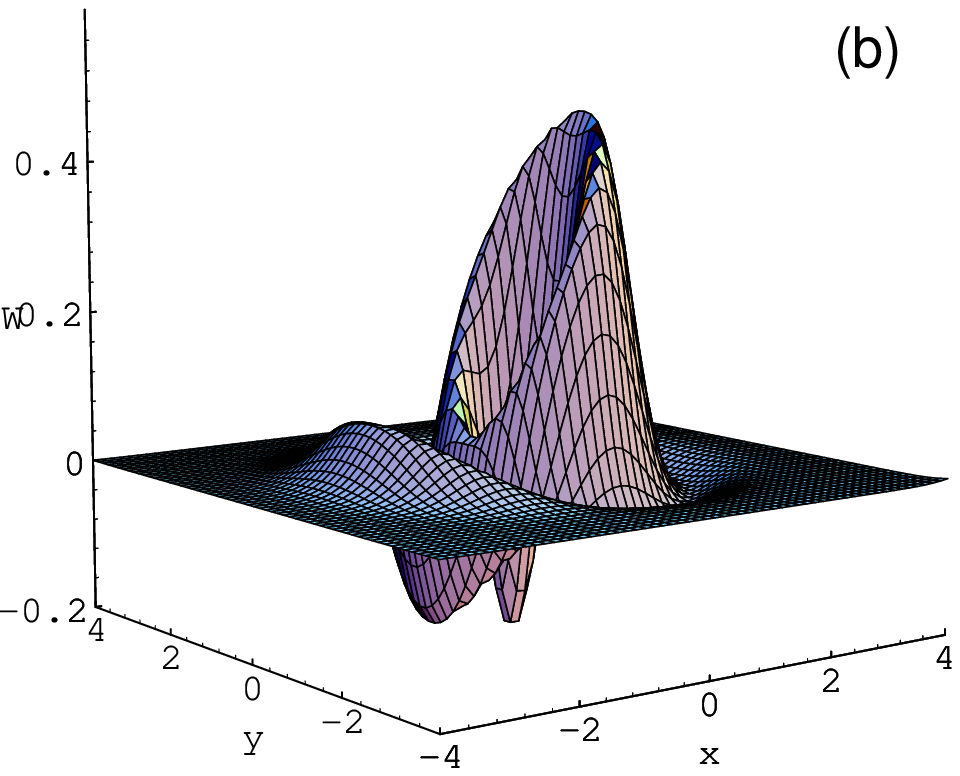}}
\centerline{\epsfxsize=8cm 
\epsfbox{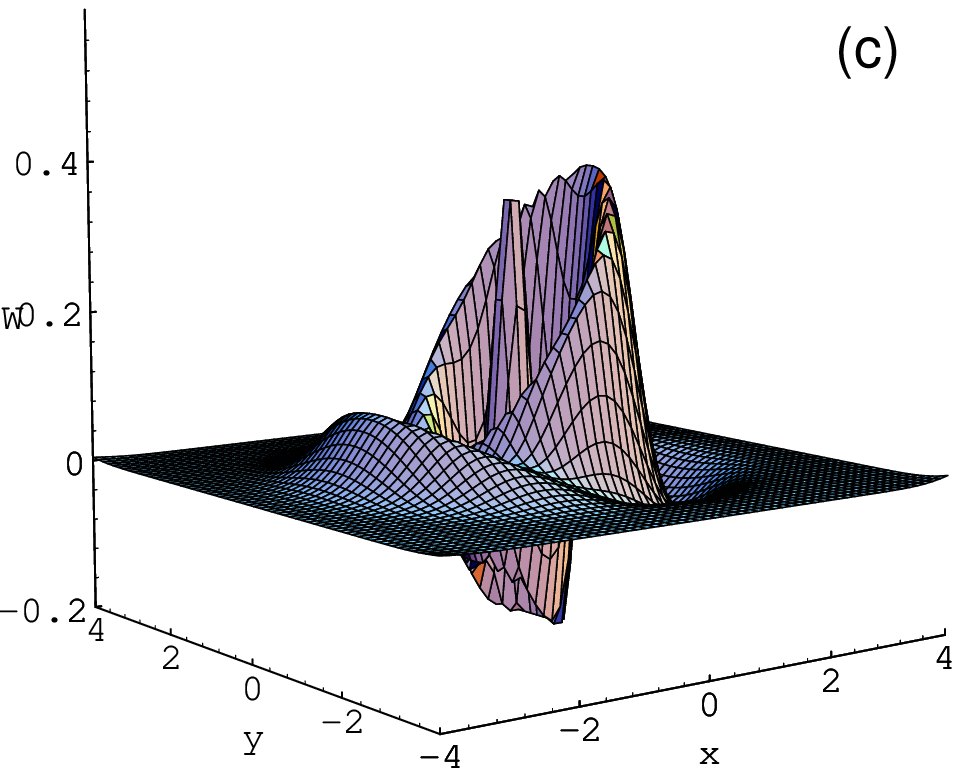}\hspace{1cm}
\epsfxsize=8cm 
\epsfbox{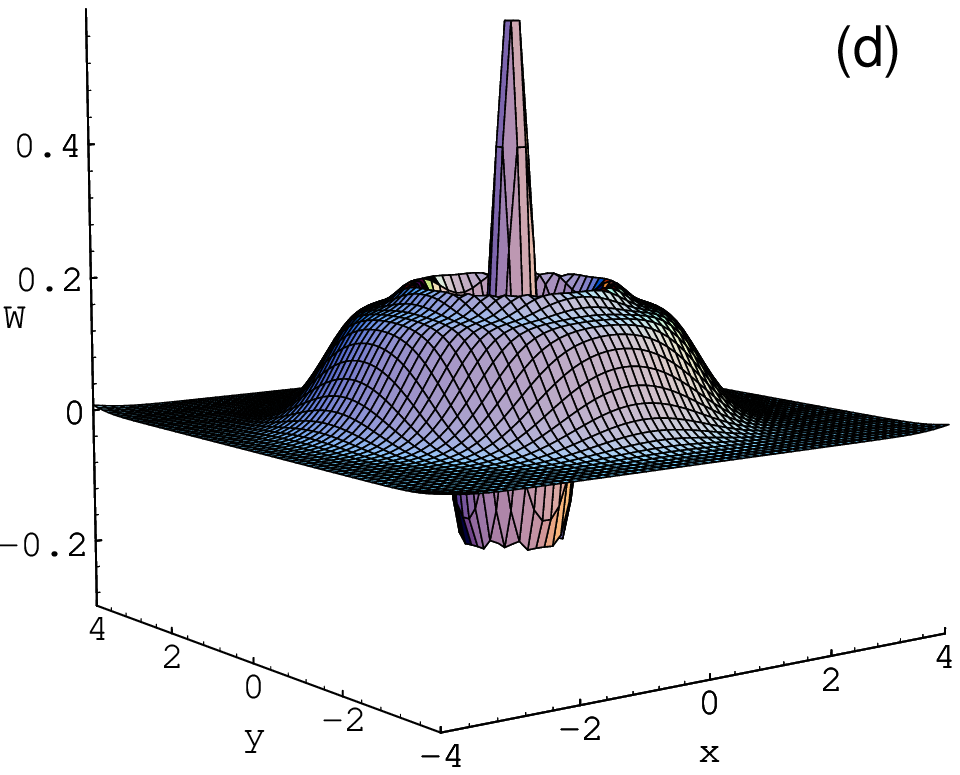}}
\caption{$\beta=x+iy$.\  Wigner function $W(\beta)$ of HGS 
for $M=2$ and different $\eta$ and $L$: (a) $\eta=0.2$, (b) 
$\eta=0.5$, (c) $\eta=0.9$ and (d) $\eta=1$ (the number 
state $|2\rangle$). The smallest possible value of $L$ is 
chosen for each $\eta$ to show the maximal  contrast with 
BS.}
\end{figure}

\begin{figure}
\centerline{\epsfxsize=8cm 
\epsfbox{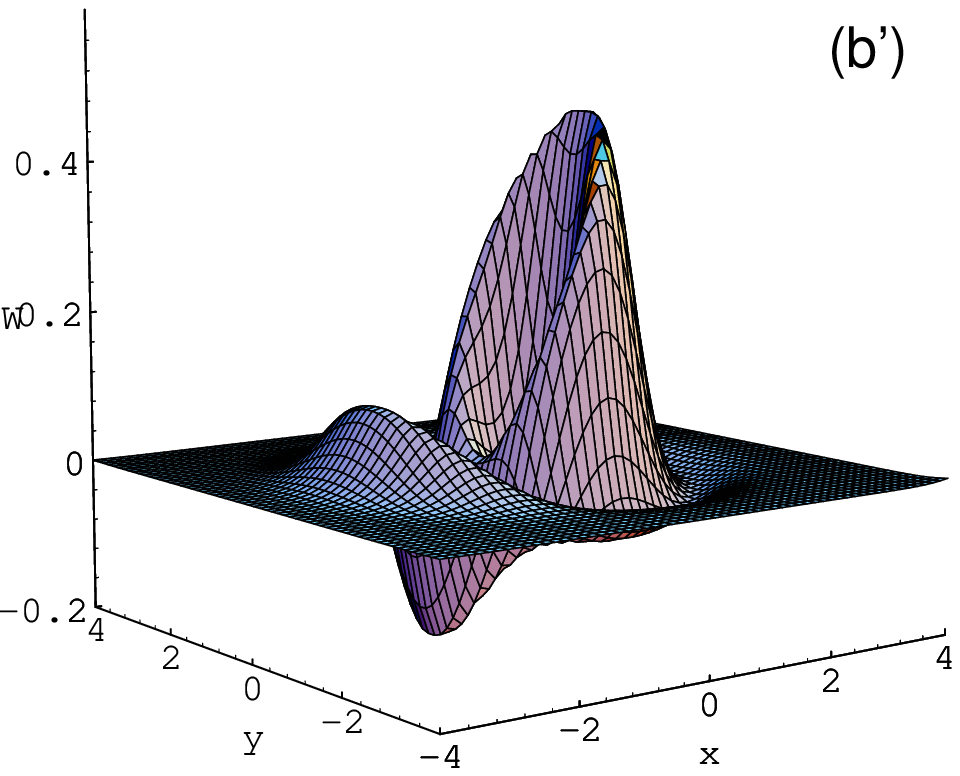}\hspace{1cm}
\epsfxsize=8cm 
\epsfbox{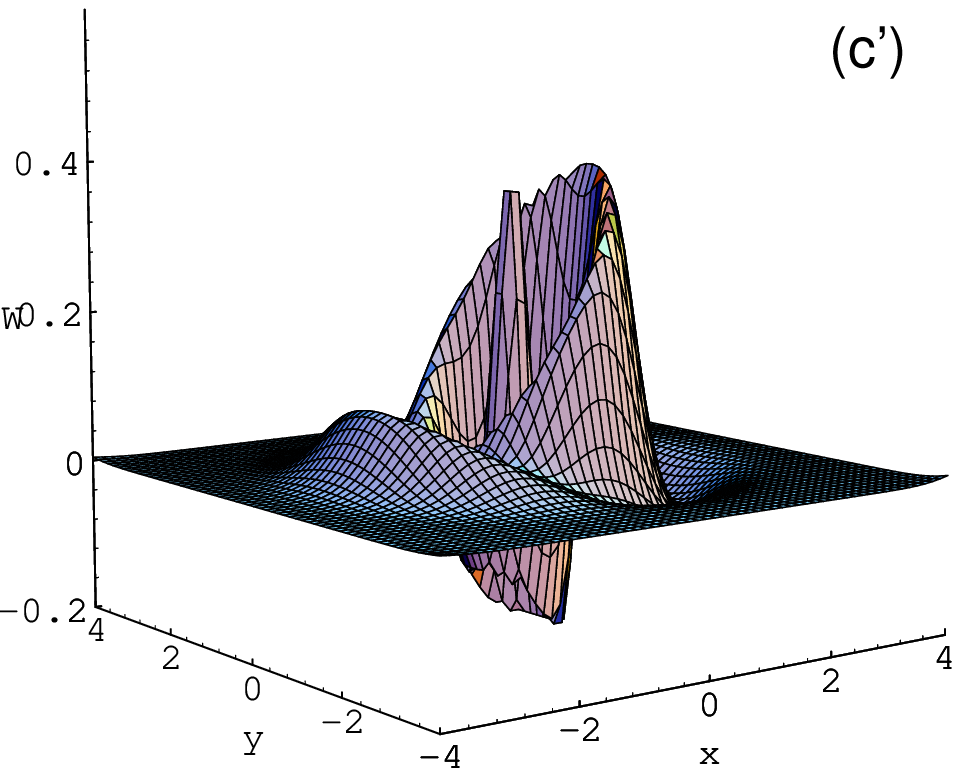}}
\caption{ Wigner function of BS for $M=2$ and (${\rm b}^\prime$) 
$\eta=0.5$ and (${\rm c}^\prime$) $\eta=0.9$.}
\end{figure}


\begin{thebibliography}{99}
\bibitem{noch}    R.\,Loudon, {\it The quantum theory of
                  light} (Clarendon Press, Oxford, 1973);   \\
                  J.\,R.\,Klauder and B.\,S.\,Skagerstam, {\it Coherent
                  states--Applications in Physics and mathematical
                  Physics} (World Scientific, Singapore, 1985);     \\
                  W.-M.\,Zhang, D.\,H.\,Feng and R.\,Gilmore,
                  Rev.\,Mod.\,Phys.\, {\bf 62}, 867 (1990).
\bibitem{stol}    D.\,Stoler, B.\,E.\,A.\,Saleh and M.\,C.\,Teich,
                  Opt.\,Acta. {\bf 32}, 345 (1985).
\bibitem{leee}    C.\,T.\,Lee, Phys.\,Rev.\,{\bf 31A}, 1213 (1985).
\bibitem{barr}    A.\,V.\,Barranco and J.\,Roversi,
                  Phys.\,Rev. {\bf 50A}, 5233 (1994).
\bibitem{datt}    G.\,Dattoli, J.\,Gallardo and A.\,Torre,
                  J.\,Opt.\,Soc.\,Am.\, {\bf 2B}, 185 (1987).
\bibitem{josh}    A.\,Joshi and R.\,R.\,Puri,
                  J.\,Mod.\,Opt. {\bf 36}, 557 (1989);  \\
                  M.\,E.\,Moggin, M.\,P.\,Sharma and A.\,Gavrielides,
                  {\it ibid.} {\bf 37}, 99 (1990).
\bibitem{Feller}  W.\, Feller, {\it An Introduction to 
                  Probability:\,Theory and Its Applications Vol.1}, 
                  (John Wiley, 1957), 2nd ed.
\bibitem{bas1}    B.\,Baseia, A.\,F.\,de Lima and A.\,J.\,da Silva,
                  Mod.\,Phys.\,Lett. {\bf 9B}, 1673 (1995).
\bibitem{bas2}    B.\,Baseia,  A.\,F.\,de Lima and G.\,C.\,Marques,
                  Phys.\,Lett. {\bf 204A}, 1 (1995).
\bibitem{fann}    H.\,Y.\,Fan and S.\,C.\,Jing, 
                  Phys.\,Rev. {\bf 50A}, 1909 (1994).
\bibitem{fus4}    H.\,C.\,Fu and R.\,Sasaki, 
                  J.\,Phys.\,{\bf 29A}, 5637 (1996) (quant-ph/9607012).
\bibitem{gdo}     D.\,Bonatsos and C.\,Daskaloyannis, Phys.\,Lett.\,
                  {\bf 307B}, 100 (1993) (and references therein); \\
                  H.\,C.\,Fu and R.\,Sasaki, J.\,Phys.\,
                  {\bf 29A}, 4049 (1996).
\bibitem{Man}     L.\,Mandel, Opt.\,Lett. {\bf 4}, 205 (1979).                 
\bibitem{WaMi}    See for example, D.\,F.\,Walls and G.\,C.\,Milburn, 
                  {\it Quantum Optics}  (Springer 1994).
\bibitem{HiOc}    M.\,Hillery, R.\,F.\,O'Connel, M.\,O.\,Scully and 
                  E.\,P.\,Wigner, Phys.\,Rep.\,{\bf 106}, 121 (1984).  
\bibitem{wigser}  H.\,Moya-Cessa and P.\,L.\,Knight, Phys.\,Rev.\,
                  {\bf 28A}, 2479 (1993).
\bibitem{1333}    V.\,Buzek, A.\,Vidiella-Barranco and P.\,L.\,Knight,
                  Phys.\,Rev.\,{\bf 45A}, 6570 (1992).
\bibitem{KenSt}   M.\,G.\,Kendall and A.\,Stuart, {\it The Advanced 
                  Theory of Statistics} (Charles Griffin Co., 1969).    
\end{thebibliography}
\end{document}